\documentclass[12pt]{article}
\pdfoutput=1 

\usepackage{hep-paper}

\renewbibmacro*{doi+eprint+url}{\usebibmacro{hep-doi+eprint+url}}
\usepackage{scalerel}							
\usepackage{subdepth}							
\usepackage{ulem}
\usepackage{subfiles}
\usepackage{import}
\usepackage{appendix}							
\AtBeginEnvironment{subappendices}{%
  \addtocontents{toc}{\protect\addvspace{10pt}Appendices}
}

\usepackage[british]{babel}						
\usepackage{upgreek}
\usepackage{xcolor}
\usepackage{verbatim}

\usepackage{tabularx}							%
\usepackage{makecell}							%
\usepackage[export = true]{adjustbox}				%

\usepackage{dsfont}

\usepackage{bibentry}
\usepackage[nottoc]{tocbibind}

\allowdisplaybreaks[1] 							

\usepackage{tocloft}

\AtBeginEnvironment{tabular}{\tlstyle}

\newcommand{\makediff}[2]{\expandafter\NewDocumentCommand\csname#1\endcsname{ogd()}{\IfNoValueTF{##2}{\IfNoValueTF{##3}{#2\IfNoValueTF{##1}{}{^{##1}}}{\mathinner{#2\IfNoValueTF{##1}{}{^{##1}}\argopen(##3\argclose)}}}{\mathinner{#2\IfNoValueTF{##1}{}{^{##1}}##2} \IfNoValueTF{##3}{}{(##3)}}}}
\makediff{mD}{\mathcal D}

\mathdef{\U}{\operatorname{U}}
\mathdef{\P}{\operatorname{P}}

\makeatletter
\renewbibmacro*{author}{\usebibmacro{hep@bib@author}}
\makeatother

\acronym{QFT}{quantum field theory}[quantum field theories]
\acronym{QCD}{quantum chromodynamics}
\acronym{QED}{quantum electrodynamics}
\acronym{GD}{gluon dynamics}
\acronym{SM}{Standard Model}
\acronym{GR}{General Relativity}
\acronym{RGE}{renormalization group equation}
\acronym{VEV}{vacuum expectation value}
\acronym{CKM}{Cabibbo-\allowbreak Kobayashi-\allowbreak Maskawa}
\acronym{PMNS}{Pontecorvo–\allowbreak Maki–\allowbreak Nakagawa–\allowbreak Sakata}
\acronym[$\chi$PT]{cPT}{chiral perturbation theory}[chiral perturbation theories]
\acronym{NGB}{Nambu-\allowbreak Goldstone boson}
\acronym{PNGB}{pseudo Nambu-\allowbreak Goldstone boson}
\acronym{WZW}{Wess-\allowbreak Zumino-\allowbreak Witten}
\acronym{MC}{Maurer-\allowbreak Cartan}
\acronym{GM}{Gell-\allowbreak Mann}
\acronym{CS}{Chern-\allowbreak Simons}
\acronym{YM}{Yang-\allowbreak Mills}
\acronym{HNL}{heavy neutral lepton}
\acronym{ALP}{axion-like particle}
\acronym{LLP}{long-lived particle}
\acronym{EOM}{equation of motion}
\acronym{PI}{partial integration}
\acronym[1PI]{onePI}{one particle irreducible}
\longacronym{WIMP}{weakly interacting massive particle}
\acronym{DM}{dark matter}
\acronym{UV}{ultraviolet}
\acronym{IR}{infrared}
\acronym{LEC}{low energy coefficient}
\acronym{LER}{low energy realisation}
\acronym{NP}{new physics}
\acronym{EM}{electromagnetic}
\acronym{EW}{electroweak}
\acronym{EWSB}{electroweak symmetry breaking}
\acronym[\overline{MS}]{MSb}{modified minimal subtraction}
\acronym{NDA}{naive dimensional analysis}
\acronym{DOF}{degree of freedom}[degrees of freedom]
\acronym{LO}{leading order}
\acronym{NLO}{next-to-leading order}
\acronym{NNLO}{next-to-next-to-leading order}
\acronym{LHC}{Large Hadron Collider}
\acronym{HL}{High Luminosity}
\shortacronym{CMS}{compact muon solenoid}
\shortacronym{ATLAS}{a toroidal LHC apparatus}
\shortacronym{LHCb}{LHC beauty}
\shortacronym{NA}*{North Area experiment \textnumero~}
\shortacronym{KOTO}{K0 at Tokai}
\acronym{SHiP}{search for hidden particles}
\shortacronym{MATHUSLA}{Massive Timing Hodoscope for Ultra-Stable neutraL pArticles}
\shortacronym{FASER}{ForwArd Search ExpeRiment}
\shortacronym[CODEX-b]{CODEXb}{compact detector for exotics at LHCb}
\shortacronym{CP}{charge-parity}
\acronym{PET}{\emph{portal effective theory}}[\emph{portal effective theories}]
\acronym{EFT}{effective field theory}[effective field theories]
\acronym{SMEFT}{Standard Model effective field theory}[Standard Model effective field theories]
\acronym{HEFT}{Higgs effective field theory}[Higgs effective field theories]
\acronym{LEFT}{light effective field theory}[light effective field theories]
\acronym{SCET}{soft-collinear effective theory}[soft-collinear effective theories]
\acronym{HQET}{heavy quark effective theory}[heavy quark effective theories]
\acronym{NRQCD}{non-relativistic quantum chromodynamics}
\acronym{NRQED}{non-relativistic quantum electrodynamics}
\acronym[2HDM]{THDM}{two Higgs-doublet model}
\acronym{SUSY}{supersymmetry}
\acronym{SUGRA}{supergravity}
\acronym{MSSM}{minimal supersymemtric Standard Model}
\acronym{NMSSM}{next-to-minimal supersymemtric Standard Model}
\longacronym{CC}{charged current}
\longacronym{NC}{neutral current}
\longacronym{LE}{low energy}[low energies]
\longacronym{HE}{high energy}[high energies]
\acronym{BR}{branching ratio}
\acronym{GW}{gravitational wave}
\acronym{QGP}{quark gluon plasma}
\acronym{GWB}{gravitational wave background}
\acronym{LISA}{Laser Interferometer Space Antenna}
\acronym{ET}{Einstein Telescope}
\acronym{DECIGO}{Deci-hertz Interferometer Gravitational wave Observatory}
\acronym{BBO}{Big Bang Observer}
\acronym{CMB}{cosmic microwave background}
\acronym{FCC}{Future Circular Collider}
\acronym{DUNE}{Deep Underground Neutrino Experiment}
\acronym{HyperK}{Hyper-Kamiokande}
\acronym{CEPC}{Circular Electron Positron Collider}
\acronym{SPPC}{Super Proton-Proton Collider}
\acronym{ILC}{International Linear Collider}
\acronym{RF}{radiofrequency}
\acronym{SR}{synchrotron radiation}
\acronym{IP}{interaction point}

\acronyms{Omit}

\acronym{SNF}{Schweizerischer Nationalfonds zur Förderung der wissenschaftlichen Forschung}
\acronym[exp]{ex}{experimental}
\acronym{lat}{lattice}
\acronym{BAU}{baryon asymmetry of the universe}
\acronym{UCLouvain}{Université catholique de Louvain}
\acronym{FSR}{Fonds Spéciaux de Recherche}
\acronym{LEP}{Large Electron-Positron Collider}
\acronym{HIKE}{High Intensity Kaon Experiment}
\acronym{BSM}{beyond the \SM}
\acronym{SSC}{Superconducting Super Collider}
\crefname{type}{type}{types}
\crefalias{inlinelisti}{type}
\crefalias{enumi}{type}

\crefformat{footnote}{#2\footnotemark[#1]#3}

\eqcrefname{lag}{Lagrangian}
\eqcrefname{set}{stress-energy tensor}
\eqcrefname{cur}{current}
\eqcrefname{sym}{symmetry}
\eqcrefname{bilinear}{quark bilinear}

\usepackage{tocloft}

\AtBeginEnvironment{tabular}{\tlstyle}

\let\oldfigure\figure
\def\figure{\oldfigure\small}
\let\oldtable\table
\def\table{\oldtable\small}

\makeatletter

\def\custom#1{{\hbox{$\left#1\vbox to30\p@{}\right.\n@space$}}}
\makeatother

\def\int{\intop\nolimits}

\bibliography{bibliography}

\title{A Possible Future Use of the LHC Tunnel\thanks{Contribution prepared for the 2025 update of the European Strategy for Particle Physics.}}

\author[louvain]{Marco Drewes\footnote{\texttt{marco.drewes@uclouvain.be}}}
\author[CERN]{Elena Shaposhnikova\footnote{\texttt{elena.chapochnikova@cern.ch}}}
\author[EPFL]{Mikhail Shaposhnikov\footnote{\texttt{mikhail.shaposhnikov@epfl.ch}}}

\affiliation[louvain]{Centre for Cosmology, Particle Physics, and Phenomenology, Université catholique de Louvain, Louvain-la-Neuve B-1348, Belgium}
\affiliation[CERN]{CERN, CH-1211 Geneva 23, Switzerland}
\affiliation[EPFL]{Institute of Physics, Ecole Polytechnique Federale de Lausanne, CH-1015 Lausanne, Switzerland}

\begin{document}

\maketitle

\begin{abstract}
The FCC program at CERN provides an attractive all-in-one solution to address many of the key questions in particle physics. While we fully support the efforts towards this ambitious path, we believe that it is important to prepare a mitigation strategy in case the program faces unexpected obstacles for geopolitical or other reasons. This approach could be based on two components: I) a circular electron-positron collider in the LHC tunnel that operates at the Z-pole energy of $45.6$~GeV and II) a high-energy electron-positron linear collider which acts as a Higgs, top quark and W-boson factory, and that can further be extended to TeV energies. The former could reach a high luminosity that is not accessible at a linear collider, the latter could probe the high energy regime with higher sensitivity and discovery potential than LEP3. The program should be flanked by dedicated intensity frontier searches at lower energies. These accelerators can be used in a feasible, timely and cost-efficient way to search for new physics and  make precise determination of the parameters of the Standard Model. 
\end{abstract}
\clearpage

\section{Particle physics at the crossroad}
The quest for a quantitative understanding of the  laws of nature has been a driving force behind the development of the natural sciences and, in particular, established the disciplines of fundamental physics and cosmology. Particle physics has played a key role in this endeavour since the second half of the 20th century. On the experimental side, the progress has largely been driven by development of modern particle accelerators leading to the establishment of 
the \SM of particle physics as the currently most fundamental description of the laws of nature at an elementary level.  The Higgs boson, the last elementary particle predicted by the \SM,  has been found experimentally at the \LHC \cite{ATLAS:2012yve,CMS:2012qbp}, and is viewed by many as the keystone of the model. 

Hence, we find ourselves in a privileged situation of possessing a theoretical framework that can describe almost all phenomena observed in nature at a fundamental level -- ranging from the inner structure of the proton to the horizon of the observable universe -- in terms of a few symmetry principles and a handful of numbers, the constants of nature.\footnote{For the sake of brevity we include classical gravity in the definition of the \SM.}
It is intriguing that the \SM is theoretically fully consistent and could, in principle, be a valid description of nature for energies up to the Planck scale $M_{P}\sim 10^{19}$ GeV \cite{Bezrukov:2012sa,Degrassi:2012ry,Bednyakov:2015sca}.

The existence of new particles with masses well below $M_P$ is nevertheless strongly motivated by two conceptually different types of arguments.
On one hand, the theory fails to describe 
several observed phenomena.
Experimental observations established beyond doubt 
are\footnote{Sometimes the accelerated expansion of the universe is included in this list, but current observations are
compatible with vacuum energy being the driving mechanism, which would be consistent with both quantum field theory and \GR.} 
\begin{inlinelist}[label = \arabic*)]
\item neutrino flavour oscillations 
indicating that neutrinos have masses 
\item the evidence for \DM 
\item the excess of matter over antimatter known as \BAU 
\item the overall geometry of the universe, including e.g., the flatness and horizon problems 
\end{inlinelist}
In addition to these four undeniable facts, there are some experimental anomalies that may point to new physics,
see, e.g.~\cite{Abdalla:2022yfr,Crivellin:2023zui}. 
Theories that can explain these phenomena within the well-established framework of \QFT generally require new fundamental fields and, thus, new elementary particles.

On the other hand, there are various aspects of the \SM  that have given rise to theoretical concerns: 
\begin{inlinelist}[label = \roman*)]
\item the values of model parameters remain unexplained,  
in particular, the large hierarchies between the 
\EW 
scale $\sim 100$ GeV, $M_P$, and the value of the cosmological constant
\item the absence of CP-violation in the strong interactions
\item the values of the Yukawa couplings, including both the hierarchy of the resulting fermion masses and the very different structures of the quark and lepton mixing matrices
\item the choice of the gauge group, including the question whether there is grand unification at high energies
\end{inlinelist}
While none of these issues strictly proves the incompleteness of the \SM in the sense of its failure to explain any observed phenomenon, they are widely believed to represent important hints towards a more fundamental theory that the \SM should be embedded in. 

To date, all attempts to find the new elementary particles 
that could solve those mysteries have failed. 
This leaves us with little experimental clues how to tackle those \emph{big questions},  
making it difficult to identify the best  strategy for the future.

\section{A Plan A for particle physics?}

The global scientific community in particle physics has identified the most important experimental and theoretical questions 
in a number of formal and informal efforts, such as the Snowmass process \cite{P5:2023wyd,Narain:2022qud,Bose:2022obr}, the European Strategy for Particle Physics \cite{EuropeanStrategyforParticlePhysicsPreparatoryGroup:2019qin} (to be updated soon), and the recent ECFA study \cite{deBlas:2024ieh}. 
An agreement on a relatively short list of key problems has been achieved, 
providing a well-defined basis for discussions of the future strategy.  

Large-scale scientific projects for the future of accelerator-based experiments have been laid out, 
including CERN's \FCC program \cite{FCC:2018byv,FCC:2018evy,FCC:2018vvp,FCC:2018bvk,2780507} and a comparably ambitious program in China \cite{Tang:2015qga,CEPCStudyGroup:2018rmc,CEPCStudyGroup:2018ghi,CEPCPhysicsStudyGroup:2022uwl,CEPCStudyGroup:2023quu}.  
We fully support these efforts, 
defining for us the \FCC program as the primary plan for particle physics in Europe.
Specifically, we endorse the strategy of building an electron collider (\FCC-ee) as the first stage, with the option to later replace it with a hadron collider (\FCC-hh) in the same tunnel,
following the approach successfully demonstrated with the \LEP and the \LHC.

\section{A Plan B for particle physics?}

While the scientific excellence of these programs is beyond doubt and, presumably, this Plan A has the support of a large fraction of high energy physicists, some risks have been pointed out in discussions within both the scientific community and  the public.  One challenge concerns the long time scales on which these projects are expected to be completed. Another potential risk lies in the availability of funding for these multi-billion euro projects.  
Given the current geopolitical situation, we believe it is advisable for the community to explore all possible avenues for discovery
and to have a Plan B in case the favourite option faces major obstacles.  

Our primary focus 
is on an experimental program that aims at the discovery of new elementary particles as well as studies of the Higgs boson and the top quark. We consider below a potential mitigation strategy, in case 
Plan A 
faces major obstacles,  with the following criteria in mind:
\begin{inlinelist}[label = \Alph*)]
\item \emph{Timeliness:} 
flagship projects 
can be realised within a time frame that 
enables discoveries as early as possible
and permits once junior scientists 
to work on 
real data before they retire  
\item \emph{Sustainability:} use the existing infrastructure as much as possible
\item \emph{Feasibility:} 
proposals should be
based on designs with studied and confirmed feasibility
\item \emph{Risk reduction:} 
implementing key parts of the program 
independently at different sites. This approach reduces reliance on individual institutions or governments while also fostering technological diversification in both development and application
\end{inlinelist}

Our proposal includes the following components
\begin{enumerate}[label = \Roman*)]
\item \label{prop:LEP3} \emph{Z-factory}: A circular electron-positron collider that primarily operates at the Z-pole, performing  precision tests of the \SM and searches for 
new elementary
particles with luminosities not achievable at a linear collider.
    \item \label{prop:ILC} \emph{Higgs/top-factory}: 
    A high-energy linear electron-positron collider that overpasses the WW production,
    acts as a Higgs factory,  can reach the $t \bar{t}$ threshold and can potentially be extended to TeV energies.
    
\end{enumerate}
This program should, of course, be complemented by other important efforts, such as experiments to search for hidden particles 
 at high intensity beam-dump facility~\cite{SHiP:2015vad,Alekhin:2015byh}, long and short baseline neutrino experiments, and precision frontier experiments (\eg using kaon, charm and beauty physics).  

For the Higgs/top factory 
a fully worked-out proposal exists, namely 
\ILC \cite{Behnke:2013xla,ILC:2013jhg,ILCInternationalDevelopmentTeam:2022izu}, which we use as a benchmark in the following.\footnote{In doing so, we do not intend to limit the range of options and sites where they may be realised, which may depend on decisions within the community and politics.}

The Z-factory 
represents the centerpiece of our proposal. It can be built in the existing \LEP/\LHC tunnel \footnote{Other, similar size tunnels would be those of UNK (see recent proposal~\cite{UNK}) or the  SSC (the tunnel to be completed), while the HERA and Tevatron tunnels are comparable to that of the CERN SPS and will be severely limited in luminosity by synchrotron radiation losses, cf.~\eg~\cite{MoortgatPick:92007}.}  and hence we shall name the project \LEP-Z throughout.

Compared to the earlier LEP${\rm 3}$ concept \cite{Blondel:1470615} (which is focused on Higgs physics), the machine proposed here would operate at a lower energy at the $Z$-pole, which enables a significantly higher instantaneous luminosity.  A Z-factory was already suggested in 1988 by C. Rubbia to exploit the increased \RF    power in  \LEP~\cite{Rubbia:1988yv}. Studies of high luminosity at this energy \cite{Jowett:223347} were aiming at $10^{32} cm^{-2} s^{-1}$, e.g.,  by increasing number of bunches per ring from $4$ to $36$.
It was suggested already in \cite{Blondel:1470615} that the LEP3 luminosity at the Z-pole can be considerably increased, above $10^{35} cm^{-2} s^{-1}$, see an example of design parameters in~\cite{Shatilov}. This is even more realistic now taking into account the latest developments, 
in particular, the new optics with crab-waist collision scheme (two apertures) and top-up beam injection achieved by installing the third (booster) ring in the tunnel, as it is foreseen for the \FCC-ee \cite{Blondel:2021ema,2780507}. 

In the following we argue that the total number of $Z$-bosons which can be produced at \LEP-Z within a reasonable time interval will be comparable to what is foreseen for \FCC-ee, \ie, a few times $10^{12}$. Indeed, one can hope that the averaged integrated luminosity per year of running  \LEP-Z will be only a factor of $\simeq 4$ lower than at \FCC-ee, due to the smaller radius of the \LEP tunnel. However, since \LEP-Z could be fully dedicated and optimised for a $Z$-pole run (while the \FCC design needs also to accommodate three higher energy accelerator configurations  in the same ring), we can expect even a better performance. Moreover, \LEP-Z can dedicate its entire lifetime to  $Z$-pole, while the running time of \FCC-ee must be shared between  $Z$-pole and the high energy runs, 
including a  high priority Higgs physics. 
These 
considerations can practically compensate for the impact of the smaller ring. Then, during a given period of time, the potential for discovery of new elementary particles and precision tests of the \SM will be comparable for \LEP-Z and the $Z$-pole runs of the \FCC-ee.
The above arguments are also valid for a comparison of LEP-Z with the original LEP3 proposal that aims to cover three physics programmes (Higgs, WW and Z). 

At the present level of the discussion
we can use scaling from the \FCC-ee luminosity at the Z-pole to corroborate our estimates.
In this case, we take the fundamental limitation of $P_{max} =  50$~Megawatts \SR power per beam (as for the \FCC-ee and the \CEPC) which determines the maximum beam current $I_b$ via the relation
    $I_b = e P_{max}/U_0$.
Here $U_0$ is the \SR loss of particle per turn
which strongly depends on the beam energy $E$ and is inversely proportional to the average bending radius of the ring $\rho$. 
The luminosity per \IP  can be presented in the form 
 $ L \propto P_{max} \rho \,\zeta_y/(E^3 \beta^*_y)$,
 with $\zeta_y$ the beam-beam parameter  and $\beta^*_y$ the beta-function (see, e.g.~\cite{Jowett:223347} for definitions of the various quantities and more details).
For a beam energy of $45.6$~GeV,
the luminosity at \LEP-Z will be smaller by the ratio of the bending radii, 
assuming the same limit for the 
$\zeta_y$ and $\beta^*_y$ 
(contributing to the transverse beam size at the collision point) at this limit. 
The value of the bending radius is in general proportional to the radius of the ring but also depends on the optics. For the latest \FCC-ee design $\rho = 10$~km.
 For the ring in the \LEP tunnel, the radius varied from $3.1$~km for \LEP to $2.6$~km for the LEP3 optics~\cite{Blondel:1470615} adopted from the LHeC design~\cite{LHeC_study}. 
 As a result the luminosity ratio for these cases would also increase from $3.2$ to $3.85$. In any case, this  scaling should be verified by rigorous design and detailed calculations.
Note, that for the present design of the \FCC-ee, at the Z-pole $U_0 = 0.039$~GeV, while it is assumed to be $0.15$~GeV for \LEP-Z~\cite{FrankZChamonix25},
and it was $0.125$~GeV in \LEP~\cite{Jowett:223347}. Finally, the new injector chain proposed~\cite{FCC:2018evy} would work for the \FCC-ee, \LEP-Z and LEP3.

Due to a very high \SR energy loss at 120 GeV, $U_0= 6.92$~GeV, LEP3 requires for the Higgs operation a high-gradient \RF  system with 8.0~GV voltage~\cite{Shatilov}. In addition, this enormous \RF  system will make a high-intensity operation at the Z-pole (requiring only \RF  voltage of 0.2~GV) very challenging, unless the staging approach is used (Z-pole comes first with a sufficient running time). 
Nevertheless LEP3 can be an option
if the other scenarios (such as Plans A and B) cannot be realised. Indeed, for Higgs production, the luminosity of the LEP3 could be comparable to that of the \ILC, see the latest review in~\cite{ILC}. However, all higher energies are excluded, (including top production), whereas studying the Higgs, WW and Z physics at the same level of accuracy would need considerably more time than in the \LEP-Z and \ILC combination.

\begin{figure}
    \centering
    \includegraphics[width=0.75\linewidth]{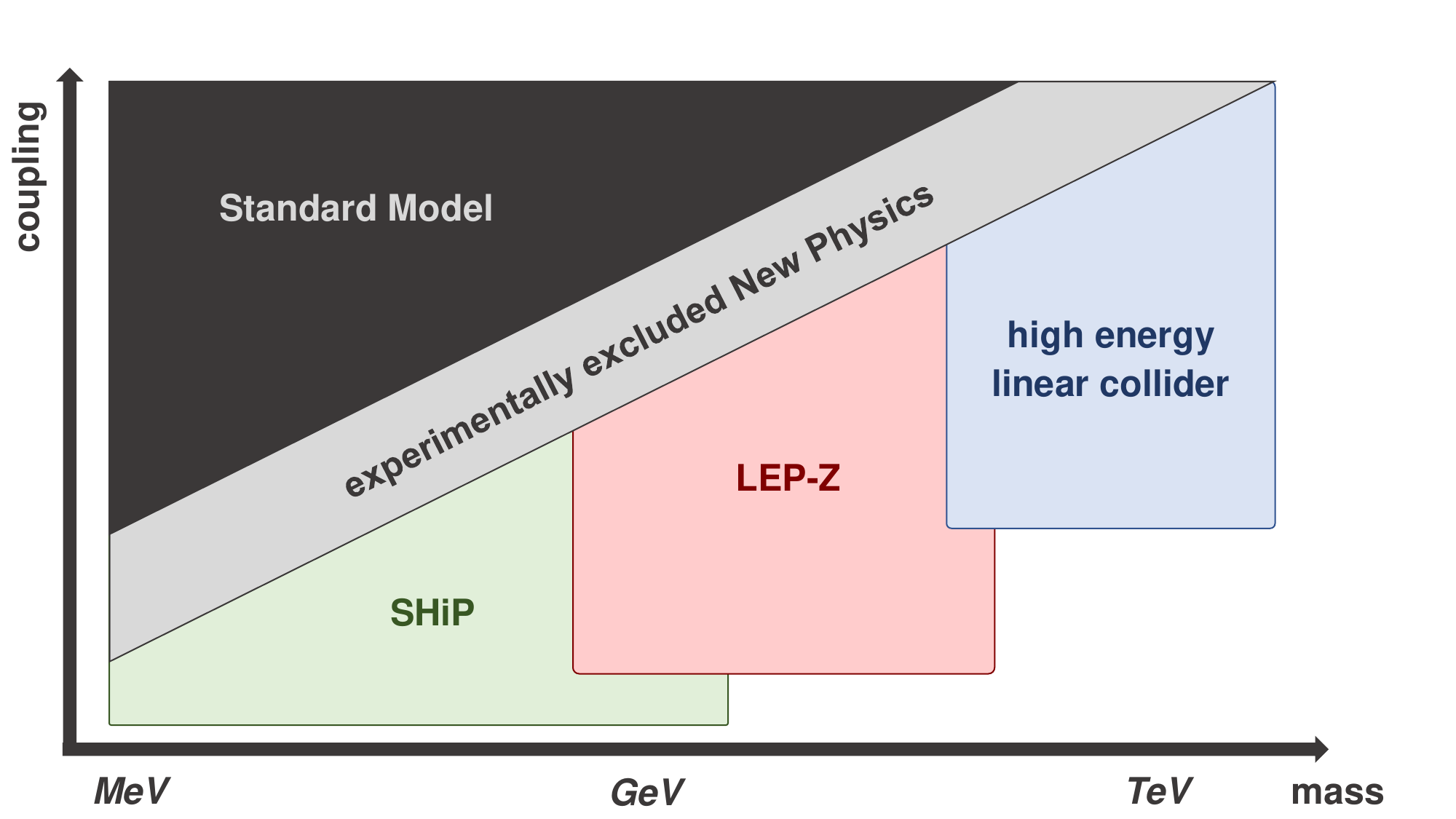}
    \caption{Schematic illustration of the complementary discovery potential of the \ILC  and \LEP-Z together with hidden sector searches. Known particles with masses below the Fermi scale and not very small couplings constitute the \SM.} 
    \label{fig:components}
\end{figure}

\section{The physics case for new accelerator facilities}\label{Sec:ExistingIdeas}

New accelerator facilities are necessitated by a number of unsolved problems in high-energy physics.
We briefly discuss here the physics case
for the 
Plan B,
using the studies for the \ILC and  the Z-pole run of the \FCC-ee.\footnote{For a discussion of the complementarity between circular and linear lepton colliders see~\cite{Blondel:2019ykp}}
Within the renormalisable \SM, the Higgs boson self-coupling is the only parameter that has not been measured,\footnote{The first generation Yukawa coupling to leptons is also only determined indirectly via the electron mass, but measuring this will be challenging even at the \FCC \cite{dEnterria:2021xij}. The unknown parameters in the neutrino sector -- mass ordering, absolute mass scale, CP-violating phases -- are strictly speaking not part of the \SM, and will in any case be targeted by dedicated neutrino experiments. } 
and would be first targeted by the linear collider. This facility can further study the properties of the top quark, 
which are e.g.~crucial to determine whether the electroweak vacuum is stable or metastable within the \SM (for a review, see \cite{Bezrukov:2014ina}).
The \ILC can produce millions of Higgs bosons and top quarks, allowing to measure the Higgs self-coupling with an uncertainty of less than $20\%$ and the top Yukawa coupling with a precision of $\sim 1\%$ \cite{ILCInternationalDevelopmentTeam:2022izu}.

  The \LEP-Z  would not only be capable of superb measurements of \EW precision observables at the  Z-pole \cite{Locci:2020knq,Bernardi:2022hny}  and flavour observables \cite{Ai:2024nmn}, but can also discover new elementary particles \cite{Blondel:2014bra,Blondel:2021ema,Blondel:2022qqo}.
The lack of their experimental observation up to now can be explained in two ways: either their mass exceeds the collision energy of the \LHC or they interact with ordinary matter so feebly that they have escaped detection due to the rarity of such events. The experimental strategy to discover them in the former case requires collisions at a higher energy (\emph{energy frontier}) 
and  more collisions in the latter case (\emph{intensity frontier}). 
It is crucial to maximise in future searches the coverage of the uncharted territory in the plane spanned by the new particles' mass and interaction strength, as illustrated in Figure~\ref{fig:components}. 

The lower left part of the figure is expected to be covered by a dedicated program to search for feebly-interacting particles based on SHiP (Search for Hidden Particles) at CERN \cite{Alekhin:2015byh,SHiP:2015vad}. This experiment, with its expected number of over $10^{20}$ protons on target, could look for very rare events, and hence probe tiny coupling constants. It would primarily produce new particles in meson decays, and hence its mass reach is limited from above by the masses of B-mesons.

Particles with masses up to the Z-mass can, if they either directly or indirectly interact with the \EW sector, be produced in gauge boson decays at \LEP-Z. In particular, searches for displaced decays can be sensitive to very small coupling constants \cite{Blondel:2022qqo}, which could be further extended with dedicated far-detectors in the caverns \cite{Wang:2019xvx,Chrzaszcz:2020emg} or on the surface \cite{Curtin:2018mvb}. 
New long-lived particles that can be found in $Z$-decays include, e.g., Heavy Neutral Leptons, axion-like particles, and additional scalar bosons 
\cite{Blondel:2014bra,Blondel:2021ema,Blondel:2022qqo}. 
An exploration of this small coupling regime shown in Figure \ref{fig:components} is not possible at a linear collider because it relies on a high luminosity. For an \ILC-like machine, the luminosity at Z-pole is expected to be smaller than that of \LEP-Z by two orders of magnitude.

At higher energies,   the search for new particles can  be taken over by a linear collider. If upgraded to  TeV energies,  
it permits, in combination with \LEP-Z and \SHiP, to explore most of the parameter space that can be covered with available technology.

\section{Risk mitigation}
The plan B offers a possible way to proceed in case the existing flagship project with all-in-one solutions in Europe
is not approved, as several potential risks may be alleviated.  
\begin{itemize}
 \item \emph{Financial risks.}  \LEP-Z can be installed in the existing tunnel at CERN after the end of HL-LHC program~\cite{HL-LHC:2020cco}, presently foreseen in $2041$.
 This would save a large amount of money on civil engineering and gives sufficient time for \LEP-Z design and preparation.  
 The construction of a linear collider can take advantage of the fact that the design of the \ILC has already been fully worked out with possible sites identified and studied. 
\item \emph{Risk of delays.}
Pursuing the physics goals with a linear collider and \LEP-Z in parallel 
minimises the impact that delays in one of these lines of research may have on the other. 
\item \emph{Technological risks.}
The linear collider and \LEP-Z  can be built with advanced but conventional technologies.
This reduces the risk of relying on technologies which still need to be developed.
\item \emph{Design optimisation.}
Running the same accelerator at  high energies and  high beam intensities  (at the Z-pole) poses  challenges for the accelerator design.
Using dedicated accelerators
for the Z-boson and Higgs/top production
ensures that each of them can be designed in an optimised way to achieve the respective physics goals.
\item \emph{Geographic diversity.} The linear collider and \LEP-Z 
can be constructed independently, one after another or 
in parallel, 
and potentially at different sites. 
This may reduce the dependence on individual funding agencies and governments. 
\item 
\emph{Physics coverage.}
Due to the complementarity of the facilities discussed here, one can simultaneously push the energy, precision,  and intensity frontiers, see Figure~\ref{fig:components}.
\item \emph{Staging possibility.}
The overall cost and resources can be leveled, minimising the peak expenditure.

\end{itemize}

\section{Conclusion}\label{Sec:Conclusion}
The existence of new elementary particles is required to explain some of the deepest mysteries about our universe.  In lack of a commonly accepted guiding principle which tells us where they are hiding, the future experimental program must aim at covering the largest possible area in the parameter space defined by the new particles' masses and couplings.  
Excellent flagship projects to achieve this goal have been outlined in both Europe and China, with  participation from the global scientific community.

In order to assure that the ambitious European experimental program in particle physics can continue even if the \FCC faces unforeseen challenges, we discuss a potential mitigation strategy (Plan B) that  allows to cover most of the uncharted territory within a foreseeable timescale and by using existing technologies.
This plan is based on two main components. The first, \LEP-Z, is a circular electron-positron collider, which operates at  energy of 45.6 GeV and acts as a Z-factory. The second component, a high energy linear electron-positron  collider (such as \ILC), acts as Higgs, top quark and W-boson factory, and can be upgraded to TeV energies. 
The \LEP-Z can achieve a high luminosity that is only lower by a factor of $\sim$4 than in \FCC-ee (and this can be compensated by a longer running time), and is not attainable at linear collider.
It could not only measure flavour and \EW precision observables at high precision, but also search for new particles with masses below the \EW scale with unprecedented discovery potential for small couplings. 
The \LEP-Z could be hosted in the existing \LHC tunnel, as the proposed LEP3. 
However, compared to LEP3, studies of the Higgs and W bosons in this approach are covered by the linear collider. 
At LEP3 these studies could only be realised, without sacrificing the sensitivity at the Z-pole, by a staging approach that significantly delays the Higgs program.
The sensitivity of the linear collider in precision test and discovery potential at high energies would exceed that of LEP3,   
and also allow for precision tests of the Higgs boson without reducing the discovery potential at the Z-pole.
When complemented by dedicated hidden sector searches for very rare events in meson decays (such as \SHiP), this program can cover most of the parameter space for new particles with masses above the MeV scale that is accessible with present day technology.

\section*{Acknowledgements}

We thank Brennan Goddard, John Jowett, Ivan Karpov, 
Fabio Maltoni, 
Frank Zimmermann, and Mikhail Zobov for helpful discussions. 

This work has been partially funded by the Deutsche Forschungsgemeinschaft (DFG, German Research Foundation) - SFB 1258 - 283604770.

\printbibliography

\end{document}